\documentclass[conference]{IEEEtran}
\IEEEoverridecommandlockouts
\usepackage{cite}
\usepackage{amsmath,amssymb,amsfonts}
\usepackage{algorithmic}
\usepackage{amsmath}
\usepackage{graphicx}
\usepackage{textcomp}
\usepackage{xcolor}
\usepackage{subfig}
\usepackage{tikz,xcolor}
\usepackage[hidelinks]{hyperref}
\usepackage{orcidlink}

\usepackage{caption, subcaption}
\def\BibTeX{{\rm B\kern-.05em{\sc i\kern-.025em b}\kern-.08em
    T\kern-.1667em\lower.7ex\hbox{E}\kern-.125emX}}
\begin{document}

\title{Reliability of Capacitive Read in Arrays of Ferroelectric Capacitors\\

\thanks{This work was supported by the European Research Council (ERC) through the European Union's Horizon Europe Research and Innovation Programme under Grant Agreement No 101042585. Views and opinions expressed are however those of the authors only and do not necessarily reflect those of the European Union or the European Research Council. Neither the European Union nor the granting authority can be held responsible for them. LF and EC would like to acknowledge the financial support of the CogniGron research center and the Ubbo Emmius Funds (Univ. of Groningen).}
}

\author{
    \IEEEauthorblockN{
        Luca Fehlings\IEEEauthorrefmark{1,3}\orcidlink{0000-0003-0993-5593},
        Muhtasim Alam Chowdhury\IEEEauthorrefmark{2} \orcidlink{0009-0008-2160-5689},
        Banafsheh Saber Latibari\IEEEauthorrefmark{2} \orcidlink{0000-0003-3735-9191}, 
        Soheil Salehi\IEEEauthorrefmark{2}\orcidlink{0000-0001-5998-8795},
        Erika Covi\IEEEauthorrefmark{1,3}\orcidlink{0000-0003-0479-6897}
    }
    \IEEEauthorblockA{
        \IEEEauthorrefmark{1}Zernike Institute for Advanced Materials \& Groningen Cognitive Systems and Materials Center (CogniGron),\\ University of Groningen, 9747 AG Groningen, The Netherlands\\ 
        \IEEEauthorrefmark{2}University of Arizona, Tucson, Arizona, USA\\
        email: \{l.d.fehlings, e.covi\}@rug.nl; \{mmc7,banafsheh,ssalehi\}@arizona.edu
    }
}

\newcommand{\luca}[1]{\textcolor{blue}{#1}}
\newcommand{\erika}[1]{\textcolor{orange}{#1}}
\newcommand{\soheil}[1]{\textcolor{red}{#1}}
\newcommand{\muhtasim}[1]{\textcolor{purple}{#1}}

\maketitle

\begin{abstract}
    The non-destructive capacitance read-out of ferroelectric capacitors (FeCaps) based on doped HfO$\mathrm{_2}$ metal-ferroelectric-metal (MFM) structures offers the potential for low-power and highly scalable crossbar arrays. This is due to a number of factors, including the selector-less design, the absence of sneak paths, the power-efficient charge-based read operation, and the reduced IR drop. Nevertheless, a reliable capacitive readout presents certain challenges, particularly in regard to device variability and the trade-off between read yield and read disturbances, which can ultimately result in bit-flips. This paper presents a digital read macro for HfO$\mathrm{_2}$ FeCaps and provides design guidelines for capacitive readout of HfO$\mathrm{_2}$ FeCaps, taking device-centric reliability and yield challenges into account. An experimentally calibrated physics-based compact model of HfO$\mathrm{_2}$ FeCaps is employed to investigate the reliability of the read-out operation of the FeCap macro through Monte Carlo simulations. Based on this analysis, we identify limitations posed by the device variability and propose potential mitigation strategies through design-technology co-optimization (DTCO) of the FeCap device characteristics and the CMOS circuit design. Finally, we examine the potential applications of the FeCap macro in the context of secure hardware. We identify potential security threats and propose strategies to enhance the robustness of the system.
\end{abstract}


\section{Introduction}
   
   The increasing volume of data in modern electronic devices is driving a shift from traditional models to a more data-centric one. With data processing speeds growing fast, conventional memory devices are falling short, particularly in scaling, reliability, and power consumption. Conventional memory architectures have a von Neumann bottleneck, but Compute-in-Memory (CiM) \cite{Backus} aims to address this issue by eliminating the architectural physical separation. This is being researched as a way of improving traditional memory systems. In recent times, ferroelectric capacitors (FeCaps), in particular those based on HfO$\mathrm{_2}$ have been scaled up to large 3D integrated memory arrays \cite{iedm23_micron} and introduced as promising efficient memory and computing devices due to their unique combination of scalability, speed, energy efficiency, and non-volatile storage \cite{10320394,8268425,10272016,yu2024ferroelectric}. 
  
   The FeCap technology exploits the distinctive property of ferroelectric materials, whereby the material's polarization can be altered by an external electric field. The devices are structured with an insulating layer comprising a ferroelectric material, such as doped HfO$\mathrm{_2}$, sandwiched between two electrodes. The key difference between FeCaps and linear capacitors is the presence of a remanent polarization, providing a non-volatile memory mechanism. While the remanent polarization is traditionally read out by a destructive reversal mechanism, recent studies on ferroelectric HfO$\mathrm{_2}$ have investigated and modeled the capacitance of metal-insulator-metal (MIM) structures \cite{kim_iedm2023, fehlings2024heracles} as a function of both the applied voltage and the polarization. This has allowed for a non-destructive readout of the polarization via the device capacitance. Highly energy-efficient and non-destructive read schemes based on this capacitance hysteresis of HfO$\mathrm{_2}$ FeCaps \cite{iedm23_capreadwindow, xiang2024compact, phadke2024reliability} have already been investigated on a device level. Moreover, operational amplifier-based circuits have been proposed for an analog readout scheme in crossbar multiply-and-accumulate applications \cite{yu2024ferroelectric, 10272016} or content-addressable memory \cite{10320394}. Nevertheless, an in-depth analysis of the reliability aspects on a memory macro level, taking into account the variability of both CMOS and FeCap technologies is still missing.
   
   Another crucial element associated with reliability is the security of the hardware. With the current advancement and complexity in the modern electronic industry, hardware security has emerged as a critical concern. Recent attacks, such as Rowhammer  \cite{6853210}, in which the data stored in nearby memory cells is altered simply by repeatedly accessing adjacent rows and causing unwanted bit-flips are critical to address for memory devices.
   Thus, ensuring integrity and reliability are critical aspects of the design, especially when sensitive data are stored or processed. Attacks such as fault injection attacks, which exploit potential vulnerabilities in the design and reliability of the hardware, are of particular concern for memory devices, whether based on traditional CMOS or emerging technologies like FeCaps and spin-orbit torque magnetic random access memories (SOT-MRAMs) \cite{ISQED2024SOT,Frontiers2024SOT}.

   The objective of this paper is to investigate the reliability of capacitive readout in a digital FeCap macro. By examining the behavior of the FeCap devices under diverse pulse voltages and read conditions, this study aims to illustrate the dependability and security implications of the FeCap memory macro in practical applications. To achieve this, we conduct a comprehensive investigation of the capacitive read reliability in the proposed FeCap memory macro, identifying the failure points and conditions that lead to read errors based on Monte Carlo simulations. According to these reliability concerns, we extend these considerations towards security aspects such as fault injection attacks.

\section{FeCap-based digital memory macro}

    \begin{figure}[t]
        \centering
        \subfloat[]{
            \includegraphics[]{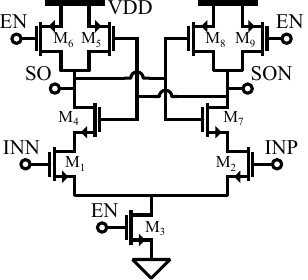}}
        \subfloat[]{
            \includegraphics[]{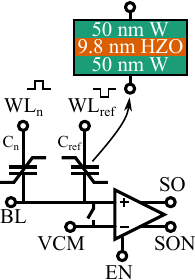}}
        \caption{(a) Sense amplifier implemented as a latch with a differential binary output signal and a differential input pair providing high input impedance. (b) Device macro with the storage FeCaps on the word lines $\mathrm{WL_{n}}$ and the reference FeCap on $\mathrm{WL_{ref}}$. For the reference capacitor a flipped FeCap is used to compensate for the asymmetric and shifted capacitance-voltage relationship. The stack of the FeCap is sketched on top.}
        \label{fig:sense-amp}
    \end{figure}
    
\begin{figure}[t]
        \centering
        \includegraphics[]{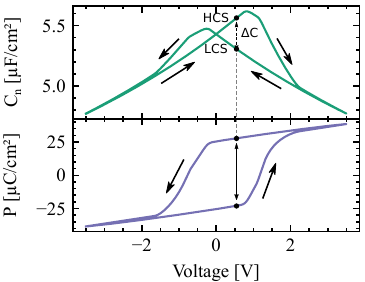}
        \caption{Capacitance (top) and polarization (bottom) hysteresis of the FeCap, simulated with a 1\,kHz triangular voltage sweep. The small signal capacitance $\mathrm{C_{FE}}$ is highest right before FeCap starts switching. The switching process, however, disturbs both the polarization state as well as the read signal and thus must be avoided. Due to the internal bias field and the non-linear C-V relationship, $\Delta$C for low read voltages can be increased by setting the polarization to intermediate states, as opposed to the saturation values.}
        \label{fig:hysteresis}
    \end{figure}

    The CMOS circuit incorporated into the memory macro is the sense amplifier (SA) shown in Fig.~\ref{fig:sense-amp}a. The SA is designed in a commercial 1.8\,V 180\,nm technology. For the purpose of simulating the FeCap, an experimentally calibrated physics-based compact model \cite{fehlings2024heracles}, Heracles, is employed. This model is accessible via the Zenodo repository \cite{heracles_zenodo}. The stack of the device is illustrated in Fig.~\ref{fig:sense-amp}b.
    The FeCap exhibits hysteresis in its capacitance (Fig.~\ref{fig:hysteresis}), which serves as an indirect mechanism for reading the polarization state. A capacitive memory window is opened, which can be utilized to differentiate between the two programmed states. In the case of Fig.~\ref{fig:hysteresis}, these are the two saturation values of the polarization, as can be observed in the polarization hysteresis (bottom graph shown in purple). The nonlinear device capacitance exhibits a higher value in the vicinity of the coercive voltage of the ferroelectric device. As the polarization of the device undergoes a transition, the capacitance corresponding to that voltage polarity shifts from the high-capacitance state (HCS) to the low-capacitance state (LCS). In addition, internal bias fields inherent to the device can result in an asymmetry in the capacitance hysteresis, thereby creating a larger capacitive memory window for one polarity, in this case for positive voltages.

    The SA (Fig.~\ref{fig:sense-amp}a) is based on a latch-type design with a high-impedance differential input pair \cite{wicht_senseamp}. The full memory macro (Fig.~\ref{fig:sense-amp}b) consists of an array with FeCaps with a reference capacitor, in addition to a transmission gate to precharge the bit line BL to the same voltage as VCM. The reference FeCap $\mathrm{C_{ref}}$ is polarized to have a capacitance in the middle of the high- and low-capacitive state, depending on the read voltage, while for the operation with a fixed read voltage it can be replaced by a conventional linear capacitor with a capacitance between HCS and LCS.
    Following Fig.~\ref{fig:timing}, the write and read operations are described. To perform a write operation, the SA is disabled by setting EN to 0\,V, so that both outputs SO and SON are high. The bit line BL and the reference capacitor $\mathrm{C_{ref}}$ are set to VCM while the programming pulse is applied to $\mathrm{WL_{n}}$. The polarization of the FeCap acting as a storage element is then set to 0\,\textmu C/cm$\mathrm{^2}$, representing the bit 1. This LCS corresponds to a net-zero polarization, between the positive and negative saturation polarization. The opposite logic state is stored by setting the FeCap to -27\,\textmu C/cm$\mathrm{^2}$, which is the HCS.
    
    The read operation of the digital FeCap macro (Fig.~\ref{fig:device-macro}) starts by pre-charging BL to VCM. Then, read pulses of 100\,mV and -100\,mV are applied to $\mathrm{WL_n}$ and $\mathrm{WL_{ref}}$, respectively. As a result, a charge $\mathrm{\Delta Q}$ is present at the high-impedance input of the SA, leading to a change in the bit line voltage relative to VCM. After pre-charging the outputs SO and SON to VDD through $\mathrm{M_{6,9}}$, the enable signal EN is set to VDD. This causes the inverter pair $\mathrm{M_{4,5,7,8}}$ to latch based on the difference in the input voltages at the differential input pair $\mathrm{M_{1,2}}$. If the stored value on the capacitor is logic 1 (logic 0), BL is higher (lower) than VCM and thus SO stays at VDD (goes to GND).
    
    The major design consideration in the FeCap macro is then to trade-off the sizing of the FeCaps and the differential input pair, here with 25\,\textmu m$\mathrm{^2}$ for the FeCap and  W\,/\,L\,=\,3\,\textmu m\,/\,0.24\,\textmu m for $\mathrm{M_{1,2}}$, to reduce the offset voltage respective to the read swing on the bit line to maximize yield while keeping a competitive speed and reducing the area overhead. Further, the read voltage can be increased for a higher signal swing, however, this comes at the cost of an increased likelihood of a read disturbance, as explained in the next Section.

    \begin{figure}[t]
        \centering
        \includegraphics[]{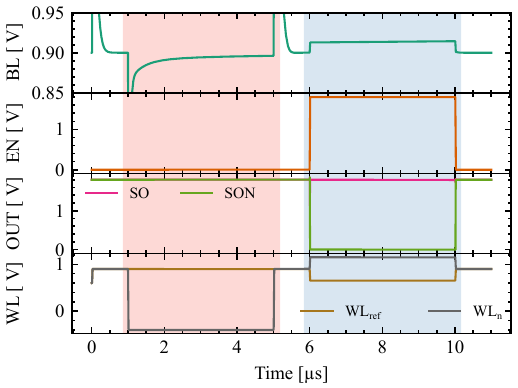}
        \caption{Timing of the programming operation (red strip) and the read operation (blue strip) of the FeCap macro. During the programming operation, the programming pulse is applied to WL while BL is shorted to VCM. The sharp voltage increases on BL are due to the displacement current originating from the programming process. During the read operation, $\mathrm{WL_{n}}$ and $\mathrm{WL_{ref}}$ receive complementary voltage pulses that charge up BL. Setting EN to VDD then triggers the outputs SO and SON to latch to the output bit and its inverted bit.}
        \label{fig:timing}
    \end{figure}

\section{Reliability and security analysis}

    \begin{figure}[t]
        \centering
        \includegraphics[]{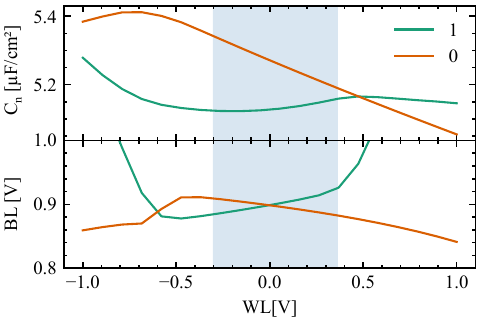}
        \caption{FeCap capacitance (top) and bit line voltage (bottom) as a function of the read voltage applied to the word line. The regime outside of the blue strip indicates where there is a substantial read disturb for the 4\,\textmu s long read pulse. The sharp increase in the BL signal for higher absolute voltages is a direct consequence of the read disturb, as the displacement current that results from the polarization switching leads to an additional signal at the bit line.}
        \label{fig:disturb_margin}
    \end{figure}
    
\begin{figure}[t]
        \centering
        \subfloat[]{
            \includegraphics[]{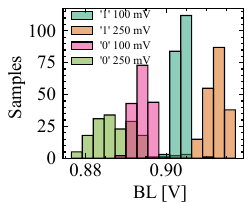}
        }
        \subfloat[]{
            \includegraphics[]{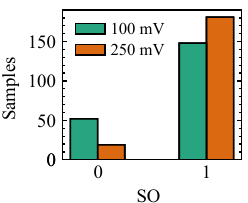}
        }
        \caption{(a) Histogram of the BL voltages for both bits and 100\,mV and 250\,mV read voltages as obtained from Monte Carlo simulations including both the programming and read operations. For the higher read voltage, the distribution of the BL voltages shifts away from the threshold, leading to an increased margin between the read signal and the SA threshold. (b) Yield of the circuit for the bit 1 written to the device. Higher input voltages lead to a higher yield as the signal at the bit line increases. However, this is at the cost of a potential read disturbance for repeated read operations.}
        \label{fig:device-macro}
    \end{figure}

    All simulations were executed at 27\,°C. The Monte Carlo simulations consist of 200 samples each and were executed with the standard parameters from the CMOS PDK and the FeCap model.
    An exploration of possible operation regimes for the read voltage is illustrated in Fig.~\ref{fig:disturb_margin}. Due to the intrinsic non-linearity of the capacitance, together with internal bias fields such as the depolarization field, the device capacitance is a function of the applied voltage. The resulting BL voltage then increases with increasing read voltage. For very high read voltages, however, a sharp rise in the BL voltage can be seen. This is due to the read process disturbing the polarization state of the FeCap, leading to a polarization current that leads to a signal at the cost of partially erasing the state, and thus causing a diminished read signal for subsequent read operations. Hence, the read voltage amplitude has to be chosen based on the specific device characteristics to trade off the read signal and yield against read disturbance and memory retention.    
    An example is shown in Fig.~\ref{fig:device-macro}, where two different read voltages are applied to read a FeCap set to logic ``1''. Monte Carlo simulations were conducted, including the variation in both programming and reading operations, and the distribution of the BL voltages is summarized in Fig.~\ref{fig:device-macro}a. The SA threshold voltage is 900\,mV with a standard deviation of 7\,mV in the Monte Carlo simulations, therefore BL voltages below this value result in a read error. A read voltage of 100\,mV results in BL voltages very close to the threshold, which leads to read errors with a probability of 25\% (Fig.~\ref{fig:device-macro}b). For a read voltage of 250\,mV the mean of the BL voltage moves further from the threshold voltage of the SA, which leads to a higher yield in the readout. This trend continues as elaborated previously until the read pulse starts leading to a substantial read disturbance which also disturbs the read signal via the polarization displacement current.

    As mentioned previously and illustrated in Fig.~\ref{fig:disturb_margin}, there is an increased likelihood of read disturb for higher read voltages. An analysis of the read disturb (Fig.~\ref{fig:disturb_capacitance}) quantifies the already mentioned trade-off. For read voltages over 200\,mV the capacitance value of the bit 1 drifts towards the reference capacitance value due to an incremental decrease of the polarization with every read operation. Thus, while higher read voltage amplitudes increase the signal, they would require a costly refresh operation. The Bit `1' is the most susceptible to read disturb in this case, as it is programmed with a negative voltage polarity and read with a positive voltage polarity, while the Bit `0' is programmed and read with a positive voltage polarity, hence the read operation reinforces the programmed state for the Bit `0'. 
    The intrinsic accumulative switching property of the FeCap is a desirable feature that enables analog computing. On the other side, the same property opens up a potential security threat: An increase in the read voltage by an attacker could cause a decrease in the yield that would go unnoticed initially and only manifest after a given amount of cycles due to the analog nature of the FeCap polarization and the ability to manipulate it via accumulative switching. This fault injection attack therefore exploits the physical properties of the FeCap to maliciously alter information inside the macro. Mitigation strategies to improve the robustness to combat fault injection attacks are described in the following Section.
    
\section{Discussion}
    
\begin{figure}[t]
        \centering
        \includegraphics[]{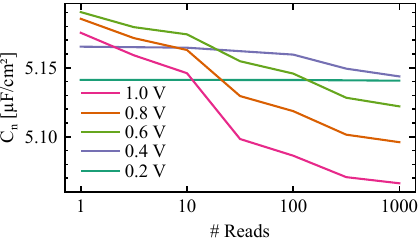}
        \caption{Capacitance of the FeCap programmed to the bit 1 over subsequent read operations with 40\,ns pulse width where the different lines represent different read voltage amplitudes. While amplitudes over 200\,mV lead to a higher signal amplitude, there is a significant read disturb that leads to the capacitance quickly dropping below the 200\,mV value and thus lead to false reads.}
        \label{fig:disturb_capacitance}
    \end{figure}

    Our findings indicate the reliability and security implications associated with the FeCap memory devices. Based on the small input signal, there is a strong incentive to optimize the read voltage amplitude to maximize the signal while minimizing the read disturbance.
    An attacker can exploit the FeCap's sensitivity to pulse voltages by launching fault injection attacks. By observing the device's behavior at different pulse voltages, adversaries can detect when the device is more prone to errors as well as small variations, as shown in Fig.~\ref{fig:disturb_capacitance} with the read disturb for increased read voltage amplitudes, and gradually and stealthily induce errors in the output data.
    
    To safeguard FeCap memory devices from such security vulnerabilities, it is critical to explore mitigation strategies. To prevent attackers from exploiting voltage-sensitive behavior in a FeCap memory macro, one approach that can be used is implementing dynamic voltage adjustment mechanisms. This will address the attacks by preventing voltage alteration in the input pulse voltage \cite{10569067}. To mitigate fault injection attacks, on-chip sensors for real-time voltage monitoring and anomaly detection can be included in the design. The on-chip sensor data paired with machine learning algorithms can assist in the timely detection and mitigation of such attacks through adjusting operations dynamically and eliminating threat vectors \cite{10558247}.
    
\section{Conclusions}
    We propose a FeCap-based non-volatile memory macro based on a digital sense amplifier that exploits the capacitance hysteresis of the FeCap to allow non-destructive reading of the device state. Based on Monte Carlo simulations for both the CMOS and the physics-based FeCap model, we show the yield achieved with a compact digital sense amplifier under different read voltages. Based on this, we discuss the reliability challenges in the design and operation process: While higher read voltages initially increase the yield, the read disturbance process becomes more pronounced due to the accumulative and partial switching commonly observed in FeCaps. We also highlight this as a potential security threat, where an attacker increasing the read voltage can cause unnoticed bit-flip errors that only manifest after a certain number of read operations, depending on the choice of the read voltage.

\section*{Acknowledgements}
The authors would like to thank Md Hanif Ali and Prof. Veeresh Deshpande for providing the electrical characterization data of the FeCap devices used to guide the design of the device macro and Dr. Fernando M. Quintana Velázquez for the fruitful discussions on the manuscript.

\bibliographystyle{ieeetr}
\bibliography{references}
\end{document}